\documentclass[11pt,reqno]{amsart}
\usepackage{amsaddr}

\usepackage{array}
\usepackage{color}
\usepackage[hidelinks]{hyperref}
\usepackage{graphicx}

\usepackage{fullpage}
\usepackage{amssymb}

\title{Solution of the time dependent Schr\"odinger equation leading to Fowler-Nordheim field emission}

\author{Ovidiu Costin}
\author{Rodica D. Costin}
\address{Department of Mathematics, The Ohio State University, Columbus, OH 43210}
\email{costin.9@osu.edu, costin.10@osu.edu}
\author{Ian Jauslin}
\address{School of Mathematics, Institute for Advanced Study}
\address{Department of Physics, Princeton University}
\email {ijauslin@princeton.edu}
\author{Joel L. Lebowitz}
\address{Departments of Mathematics and Physics, Rutgers University, Hill Center - Busch Campus, 
110 Frelinghuysen Road Piscataway, NJ 08854}
\email {lebowitz@math.rutgers.edu}

\numberwithin{equation}{section}

\begin{document}
\maketitle

\today

\begin{abstract} 
We solve the time-dependent Schr\"odinger equation describing the emission of electrons from a metal surface by an external electric field $E$, turned on at $t=0$. Starting with a wave function $\psi(x,0)$, representing a generalized eigenfunction when $E=0$, we find $\psi(x,t)$ and show that it approaches, as $t\to\infty$, the Fowler-Nordheim tunneling wavefunction $\psi_E$. The deviation of $\psi$ from $\psi_E$ decays asymptotically as a power law $t^{-\frac32}$. The time scales involved for typical metals and fields of several V/nm are of the order of femtoseconds. We plot the short-time evolution of the current and density.
\end{abstract}

\section{Introduction}
      
The emission of electrons from a cold metal surface subjected to a constant (or oscillating) electric field is a subject of great practical and theoretical interest\-~\cite{Je03,Fo16,Je17}. The microscopic theory of such emissions by a constant field was developed by Fowler and Nordheim (FN) in the early days of quantum mechanics\-~\cite{FN28} (referred to then as the ``new mechanics''). They  considered an idealized situation in which the electrons  in the conduction band are treated, a la Sommerfeld, as free independent particles. Their energies are described by a Fermi distribution with maximum energy $E_F=\hbar^2k_{\mathrm F}^2/2m$; the deviation from this zero-temperature distribution is negligible at room temperatures. In the absence of an external field the electrons are confined by an external potential (caused by the positive ions) of magnitude $U=E_F+W$, where $W$ is the work function, i.e. the energy necessary to extract an electron from the metal.
\bigskip

\begin{figure}
  \hfil\includegraphics[width=8cm]{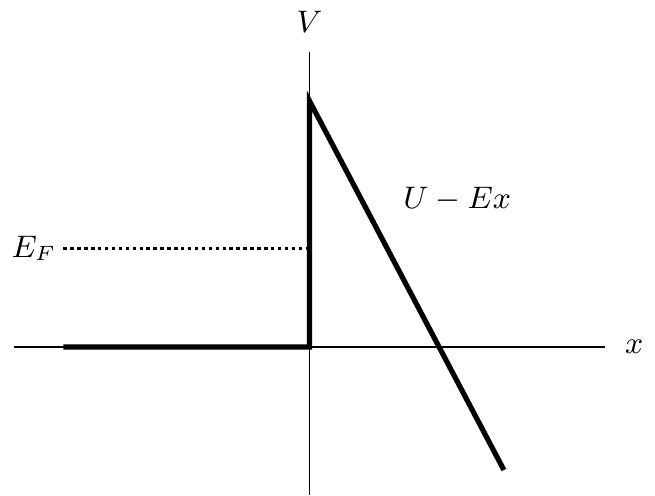}\hfil\hbox{}\par
  \caption{The shape of the potential $V(x)$.}
  \label{fig:potential}
\end{figure}

Considering emissions perpendicular to a flat surface at $x=0$, obtained when applying an external field $E$ for $x\geqslant 0$, assuming that the metal occupies all space $x<0$, leads to a one-dimensional tunneling problem in a triangular potential, see Fig.\-~\ref{fig:potential}. The one-dimensional Schr\"odinger equation describing an electron moving in this potential is then given by
\begin{equation}
  \label{schrodinger}
  i\partial_t\psi(x,t)=(-{\textstyle\frac12}\partial_x^2+V(x))\psi(x,t)
\end{equation}
(we write $\partial_x\equiv\frac\partial{\partial x}$) where
\begin{equation}
  V(x)=\left\{ \begin{array}{ll} 0, & x<0\\
  U-E x, & x>0 \end{array}\right.  
  \label{V}
\end{equation}
in atomic units $(\hbar=m=|e|=1)$.
\bigskip

When $E=0$, the potential is, simply, a step function. The Schr\"odinger equation\-~(\ref{schrodinger}) with $E=0$ has stationary solutions with energies $k^2/2<U$, $\psi(x,t)=e^{-i\frac12k^2t}\psi_0(x)$, with $k>0$ and
\begin{equation}
  \psi_0(x)=\left\{ \begin{array}{ll} e^{i k x} +R_0 e^{-i k x} & x<0\\
  T_0e^{-\sqrt{2U-k^2}x} & x>0 \end{array}\right.  
  \label{psi0}
\end{equation} 
in which $R_0$ and $T_0$ are the {\it reflection} and {\it transmission} coefficients (we use a normalization in which the amplitude of the incoming wave with $k>0$ is 1):
\begin{equation}
  R_0=\frac{ik+\sqrt{2U-k^2}}{ik-\sqrt{2U-k^2}}
  ,\quad
  T_0=\frac{2ik}{ik-\sqrt{2U-k^2}}
  .
  \label{R0T0}
\end{equation}
These constants ensure that $\psi_0(x)$ and $\partial_x\psi_0(x)$ are continuous at $x=0$. Note that, in this state, the current vanishes:
\begin{equation}
  j_0(x)=i(\psi_0\partial_x\psi_0^*-\psi_0^*\partial_x\psi_0)=0
  .
\end{equation}
\bigskip

When $E>0$, there is the possibility for an electron moving in the $+x$ direction, with kinetic energy $k^2/2<U$, to tunnel through the potential barrier and be emitted. This will then produce an electron current in the $+x$-direction. To obtain the probability of tunneling, FN computed the stationary solutions $\psi(x,t)=e^{-i\frac12k^2t}\psi_E(x)$ by solving
\begin{equation}
  (-{\textstyle\frac12}\partial_x^2+\Theta(x)(U-Ex)-{\textstyle\frac12}k^2)\psi_E(x)=0
  \label{eqpsiE}
\end{equation}
($\Theta(x)$ is the Heaviside function, which is equal to 1 if $x\geqslant 0$ and $0$ otherwise) whose solution is
\begin{equation}
  \psi_E(x)=
  \left\{ \begin{array}{ll}
    e^{i k x} +R_E e^{-i k x} & x<0\\
    T_E\Phi(x) & x>0.
  \end{array}\right.  
  ,\quad
  k>0
  \label{psiE}
\end{equation} 
in which $\Phi(x)$ is proportional to the Airy  function Ai$(x)$ (or the equivalent expression in terms of Hankel or Bessel functions), which decays when $x\to\infty$, and yet has a constant positive current for all $x$. This solution, see also\-~\cite{Ro11,Je03}, yielded the tunneling probability $D(k)=1-|R_E|^2$ of the electron as a function of $k,\,U$ and $E$. Integrating $kD(k)$ over the ``supply function'' corresponding to the density of electrons in the Fermi sea moving in the $+x$ direction with energy $k^2/2$, leads to an expression for the total steady state current $j_E$ in a static field $E$. An approximate expression for $j_E$ is~\-\cite{Fo08b,Je17}
\begin{equation}
  j_E\approx c_1E^2e^{-\frac{c_2}E}
  .
\end{equation}
\bigskip
 
The FN formula for $j_E$, with various corrections for the idealizations made, e.g. flat surface, independent electrons, neglecting the Schottky effect, {\it etc.}, serves as the backbone of cold electron emission theory and experiment.  There is a vast literature on the subject (the original FN paper\-~\cite{FN28} has more than 6000 citations). We cite here only a few\-~\cite{Ro11,Fo08} and refer the reader for more information to the recent book by Jensen\-~\cite{Je17} and references therein.
\bigskip

In this note we shall be concerned with a different problem, which, as far as we know, has not been investigated fully before. As an initial condition, we take a stationary solution of the Schr\"odinger equation at $E=0$, $\psi_0(x)$ in\-~(\ref{psi0}), and, at $t=0$, we turn the field on, and study the time evolution. In particular, we will investigate how long it will take, if ever, for the initial state $\psi(x,0)$ to approach the stationary state  $\psi_E(x)$ in\-~(\ref{psiE}). Of course, turning on $E$ instantaneously is an idealization, which we shall accept here. (In\-~\cite{YGR11}, this initial condition is considered, but the analysis then focuses mostly on the stationary solution.)
\bigskip

In what follows, we shall prove that, for $\psi(x,0)=\psi_0(x)$, $\psi (x,t)$ approaches, for long times, the $\psi_E(x)$ of\-~(\ref{psiE}), i.e.,
\begin{equation}
  \label{psi_lim}
  \psi (x,t)\sim e^{-i\frac12k^2t}\psi_E(x)
\end{equation}
In fact, this holds for a wider class of initial conditions, in which the initial incident wave is $e^{ikx}$ and the initial reflected and transmitted waves are arbitrary. The deviation $\psi(x,t)-\psi_E(x)$ decays asymptotically as $t^{-\frac32}$. The actual time dependence, of course, depends on the exact form of $\psi(x,0)$. We shall calculate this for the $\psi(x,0)=\psi_0(x)$ given in\-~(\ref{psi0}) for different values of the parameters.
\bigskip

Roughly speaking we find that for $U\approx9\ \mathrm{eV}$, $\hbar^2k_{\mathrm F}^2/2m=E_{\mathrm F}\approx 4.5\ \mathrm{eV}$ and $E\approx 4$-$8\ \mathrm{V}\cdot\mathrm{nm}^{-1}$, the time for the density $|\psi|^2$ and the current $j(t)$ to approach its final FN value is of the order of femtoseconds. The exact value depends on the position $x$ where we measure the current: for larger $x$, the time it takes for the current to stabilize is larger, see Fig.\-~\ref{current_density}. Such time scales are of practical relevance for short pulses of the order of femtoseconds or less. These are now common for oscillating laser fields for which the initial value problem will be considered in a later paper. (The ``steady state'' solution for laser fields was investigated in detail by Faisal et al\-~\cite{FKS05}; see also \cite{ZL16}.)

\section{Solution of the initial value problem}

In order to emphasize the role of each term in the initial condition, we will split $\psi(x,0)$ into three terms: an incoming, a reflected, and a transmitted wave.
\begin{equation}
  \psi(x,0)=\psi^{(\mathrm I)}(x,0)+\psi^{(\mathrm R)}(x,0)+\psi^{(\mathrm T)}(x,0)
  \label{psi_sum}
\end{equation}
with
\begin{equation}
  \psi^{(\mathrm I)}(x,0)=\Theta(-x)e^{ikx}
  ,\quad
  \psi^{(\mathrm R)}(x,0)=R_0\Theta(-x)e^{-ikx}
  ,\quad
  \psi^{(\mathrm T)}(x,0)=T_0\Theta(x)e^{-\sqrt{2U-k^2}x}
   ,\quad
   k>0
\end{equation}
(recall that $\Theta(x)$ is the Heaviside function, which is equal to 1 if $x\geqslant 0$ and $0$ otherwise). Since the Schr\"odinger equation is linear, its solution will be the sum of the solutions for each term in $\psi(x,0)$.
\bigskip

To obtain $\psi(x,t)$ we solve for $\hat\psi_p(x)$, the Laplace transform of $\psi(x,t)$,
\begin{equation}
  \hat\psi_p(x):=\int_0^\infty dt\ e^{-pt}\psi(x,t)
  \label{psip}
\end{equation}
which we obtain in closed form. We then compute, by inverting the Laplace transform, the long time asymptotics analytically, and the short time behavior numerically. This method provides an integral representation of the solution which can be evaluated numerically. It is thus better for our purposes than direct computations of the solution of\-~(\ref{schrodinger}). The latter requires cutoffs for the non-square integrable functions we are dealing with and cannot be used for long times. The Laplace transform of $\psi$ satisfies the equation
\begin{equation}
  (-{\textstyle\frac12}\partial_x^2+\Theta(x)(U-Ex)-ip)\hat\psi_p(x)=-i\psi(x,0)
  .
  \label{schrodinger_laplace}
\end{equation}
The physical solution to this equation is
\begin{equation} \label{sol11}
  \hat\psi_p(x)=
  \left\{\begin{array}{>\displaystyle ll}
    C_1(p)e^{\sqrt{-2ip}x}+F^{(\mathrm I)}_p(x)+R_0F^{(\mathrm R)}_p(x)
    &\mathrm{if\ }x<0\\[0.5cm]
    C_2(p)\varphi_p(x)+T_0F^{(\mathrm T)}_p(x)
    &\mathrm{if\ }x> 0
  \end{array}\right.
\end{equation}
where $R_0$ and $T_0$ are given in\-~(\ref{R0T0}),
\begin{equation}\label{2p6}
  F^{(\mathrm I)}_p(x):=-\frac{2ie^{ikx}}{-2ip+k^2}
  ,\quad
  F^{(\mathrm R)}_p(x):=-\frac{2ie^{-ikx}}{-2ip+k^2}
\end{equation}
\begin{equation}
  F^{(\mathrm T)}_p(x):=
  \frac{4\pi}{(2E)^{\frac13}}\left(\varphi_p(x)\int_0^x dy\ \eta_p(y)e^{-\sqrt{2U-k^2}y}
  +\eta_p(x)\int_x^\infty dy\ \varphi_p(y)e^{-\sqrt{2U-k^2}y}\right)
\end{equation}
and
\begin{equation}
  \varphi_p(x)=\mathrm{Ai}\left(2^{\frac13}e^{-\frac{i\pi}3}\left(E^{\frac13}x-E^{-\frac23}(U-ip)\right)\right)
\end{equation}
\begin{equation}
  \eta_p(x)=e^{-\frac{i\pi}3}\mathrm{Ai}\left(-2^{\frac13}\left(E^{\frac13}x-E^{-\frac23}(U-ip)\right)\right)
  \label{varphi}
\end{equation}
are two independent solutions of $(-\frac12\partial_x^2+U-Ex-ip)f=0$. The phases $e^{-\frac{i\pi}3}$ and $-1$ are cube roots of $-1$. The constants $C_1(p)$ and $C_2(p)$ are set so that $\hat\psi_p$ and $\partial_x\hat\psi_p$ are continuous at $x=0$:
\begin{equation}
  C_1(p)=-\frac{2iT_0}{\sqrt{-2ip}\varphi_p(0)-\partial\varphi_p(0)}\left(
    \frac{\sqrt{2U-k^2}\varphi_p(0)+\partial\varphi_p(0)}{-2ip+k^2}
    +\int_0^\infty dy\ \varphi_p(y)e^{-\sqrt{2U-k^2}y}
  \right)
\end{equation}
and
\begin{equation}\label{formC2}
  \begin{array}{>\displaystyle r@{\ }>\displaystyle l}
    C_2(p)=-\frac{2iT_0}{\sqrt{-2ip}\varphi_p(0)-\partial\varphi_p(0)}&\Bigg(
      \frac{\sqrt{2U-k^2}+\sqrt{-2ip}}{-2ip+k^2}
      \\\hfill&\hskip10pt
      -\frac{2i\pi}{(2E)^{\frac13}}(\sqrt{-2ip}\eta_p(0)-\partial\eta_p(0))\int_0^\infty dy\ \varphi_p(y)e^{-\sqrt{2U-k^2}y}
    \Bigg)
  \end{array}
\end{equation}
where $\partial\varphi_p(0)\equiv\frac{\partial\varphi_p(x)}{\partial x}\big|_{x=0}$ and similarly for $\partial\eta_p(0)$. The square root is defined with a branch cut along the positive imaginary axis, in such a way that $\sqrt{-2ip}$ has a branch cut along the real negative axis.
\bigskip

A simple calculation shows that, as expected,
\begin{equation}
  \lim_{\displaystyle\mathop{\scriptstyle|p|\to\infty}_{\mathcal Re(p)>0}}p\hat\psi_p(x)=\psi(x,0)
  \label{laplace_init}
\end{equation}
which confirms that $\hat\psi_p(x)$ is, indeed, the Laplace transform of a function whose initial condition is $\psi(x,0)$.
\bigskip

We then invert the Laplace transform:
\begin{equation}
  \psi(x,t)=\frac1{2i\pi}\int_{\gamma-i\infty}^{\gamma+i\infty}dp\ e^{pt}\hat\psi_p(x)
  \label{inv_laplace}
\end{equation}
in which $\gamma>0$ is an arbitrary small parameter taken close to $0$. 
\bigskip

\begin{figure}
  \hfil\includegraphics[width=8cm]{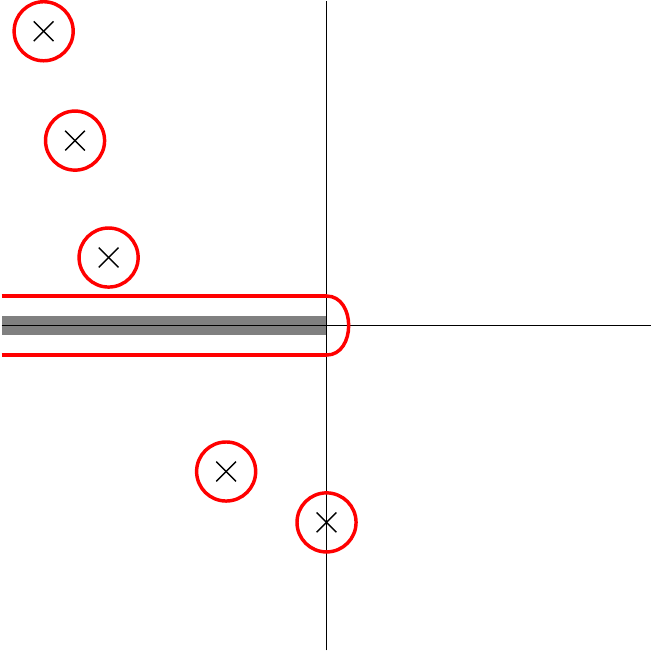}\hfil\hbox{}\par
  \caption{The deformed integration contour goes around the poles (one of which is on the imaginary axis, at $-ik^2/2$, while the others are in the negative real half-plane) and goes along the branch cut on the real negative axis.}
  \label{fig:contour}
\end{figure}

As is well known the integral on the right hand side of\-~(\ref{inv_laplace}) can be computed deforming the integration contour as in Fig.\-~\ref{fig:contour}, and studying the singularities, poles and branch points of $\hat\psi_p(x)$, lying in the half plane $\mathcal Re(p)\leq 0$. In particular, the only terms which do not decay as $t\to\infty$ come from poles on the imaginary $p$-axis. Analyzing\-~(\ref{sol11})-(\ref{formC2}) we find that the singularities of $\hat\psi_p(x)$ are, for $k>0$,
\begin{itemize}
  \item a pole on the imaginary axis, located at $-ik^2/2$, coming from  \eqref{2p6}, \eqref{formC2} and \eqref{laplace_init}
  \item  poles with strictly negative real parts corresponding to the roots of $\sqrt{-2ip}\varphi_p(0)-\partial\varphi_p(0)$ appearing in the denominators of $C_1$ and $C_2$,
  \item a branch cut along the negative real axis coming from $\sqrt{-2ip}$.
\end{itemize}
\bigskip

\subsection{Long time behavior}
The residue at $-ik^2/2$ yields the only term which does not decay in time: by an explicit computation, we find that the residue is equal to
\begin{equation}
  e^{-i\frac12k^2t}\psi_E(x)
\end{equation}
where $\psi_E$ is the FN solution\-~(\ref{psiE}).
\bigskip

The residues of the poles with a negative real part decay exponentially in time (because of the factor $e^{pt}$ in~\-(\ref{inv_laplace})).
\bigskip

The integral along the branch cut decays algebraically, as $t^{-\frac32}$: we define, for $p-i\epsilon\in\mathbb R_-$,
\begin{equation}
  \alpha:=e^{\frac{i\pi}4}\sqrt{-ip}
  ,\quad
  f(\alpha):=\hat\psi_{\alpha^2}(x)
\end{equation}
(recall the definition of $\hat\psi_p$ in~\-(\ref{psip})) and write the integral along the branch cut as
\begin{equation}
  \psi^{(\mathrm{BC})}(x,t)
  :=
  \int_{-\infty-i\epsilon}^{-i\epsilon} dp\ e^{pt}\hat\psi_p(x)
  +
  \int_{i\epsilon}^{-\infty+i\epsilon} dp\ e^{pt}\hat\psi_p(x)
  =
  2\int_0^\infty d\alpha\ e^{-\alpha^2t}\alpha(f(\alpha)-f(-\alpha))
  .
\end{equation}
By Taylor expansion, (in this context, this technique is usually called Watson's lemma)
\begin{equation}
  \psi^{(\mathrm{BC})}(x,t)
  =
  4\int_0^\infty d\alpha\ e^{-\alpha^2t}\alpha^2\partial f(0)
  +O(t^{-\frac52})
  =\left(\frac{t}{\tau_E(x)}\right)^{-\frac32}+O(t^{-\frac52})
  .
\end{equation}
with
\begin{equation}
  \tau_E(x)=\left\{\begin{array}{>\displaystyle ll}
    \left(c_E(\varphi_0(0)+x\partial\varphi_0(0))\right)^{\frac23}
    &\mathrm{if\ }x<0
    \\[0.5cm]
    \left(c_E\varphi_0(x)\right)^{\frac23}
    &\mathrm{if\ }x>0
  \end{array}\right.
\end{equation}
and
\begin{equation}
  c_E=
  -\frac{\sqrt2T_0e^{\frac{i\pi}4}}{\sqrt\pi(\partial\varphi_0(0))^2}
  \left(
    \frac{\sqrt{2U-k^2}\varphi_0(0)+\partial\varphi_0(0)}{k^2}
    +\int_0^\infty dy\ \varphi_0(y)e^{-\sqrt{2U-k^2}y}
  \right)
  .
\end{equation}

All in all, we find that
\begin{equation}
  \psi(x,t)
  =e^{-i\frac12k^2t}\psi_E(x)+\left(\frac{t}{\tau_E(x)}\right)^{-\frac32}+O(t^{-\frac52})
  .
\end{equation}
Therefore, the wave function tends to the Fowler-Nordheim solution, with a rate $t^{-\frac32}$.
\bigskip

\subsection{Short time behavior}
The behavior of $\psi(x,t)$ for small $t$ is more difficult to study analytically, but the inverse Laplace transform\-~(\ref{inv_laplace}) yields an integral formula that can be efficiently approximated numerically using fast Fourier transforms.
\bigskip

\begin{figure}
  \begin{tabular}{c|c}
    \includegraphics[width=7cm]{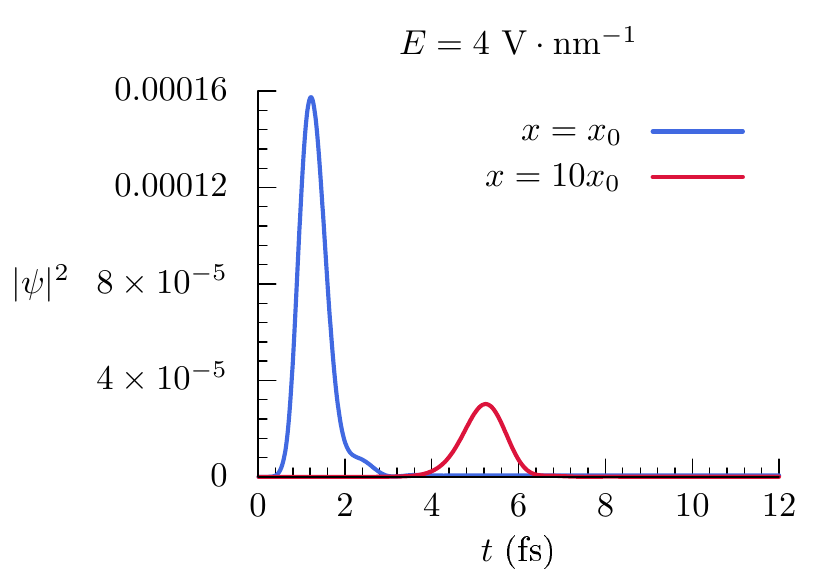}           {\scriptsize({\bf a})}&
    \includegraphics[width=7cm]{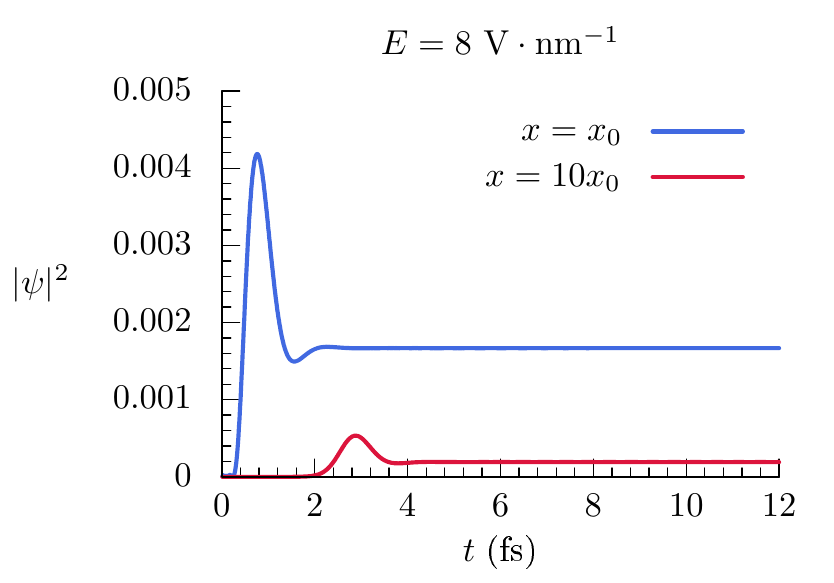}           {\scriptsize({\bf b})}\\
    \hline
    \includegraphics[width=7cm]{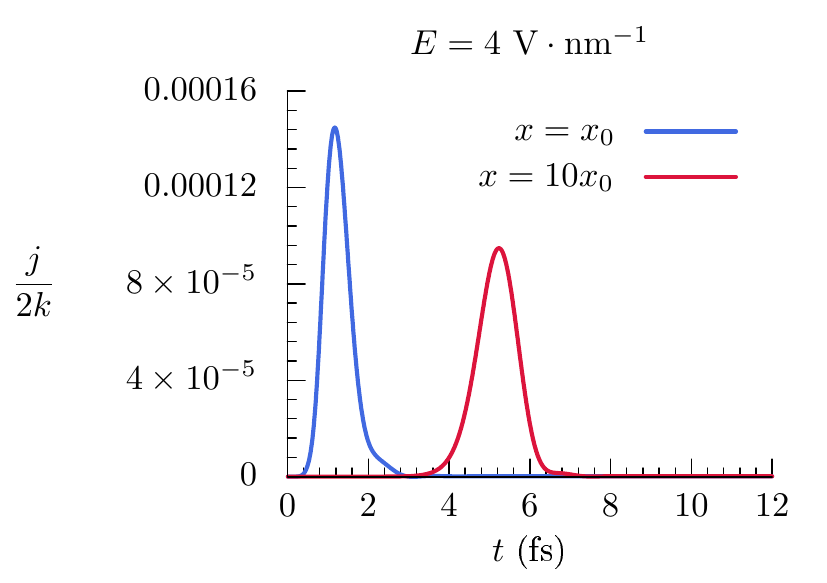}          {\scriptsize({\bf c})}&
    \includegraphics[width=7cm]{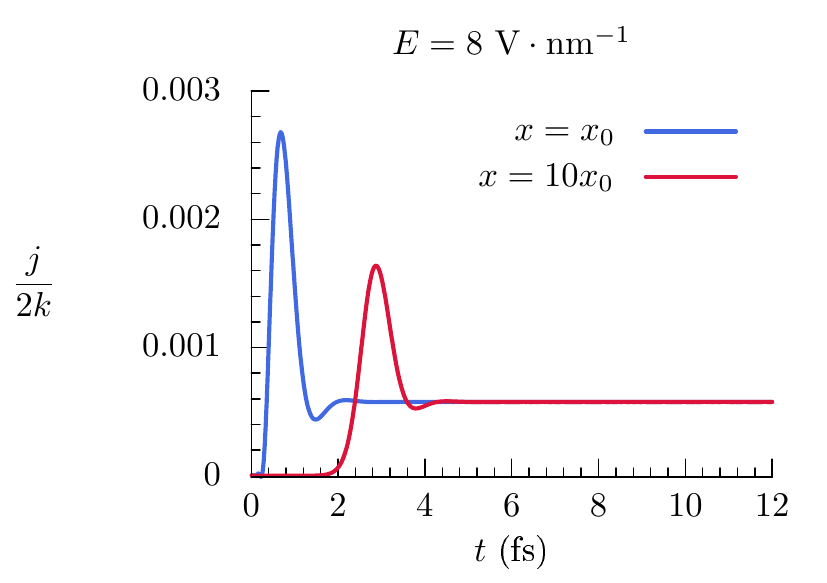}          {\scriptsize({\bf d})}\\
    \hline
    \includegraphics[width=7cm]{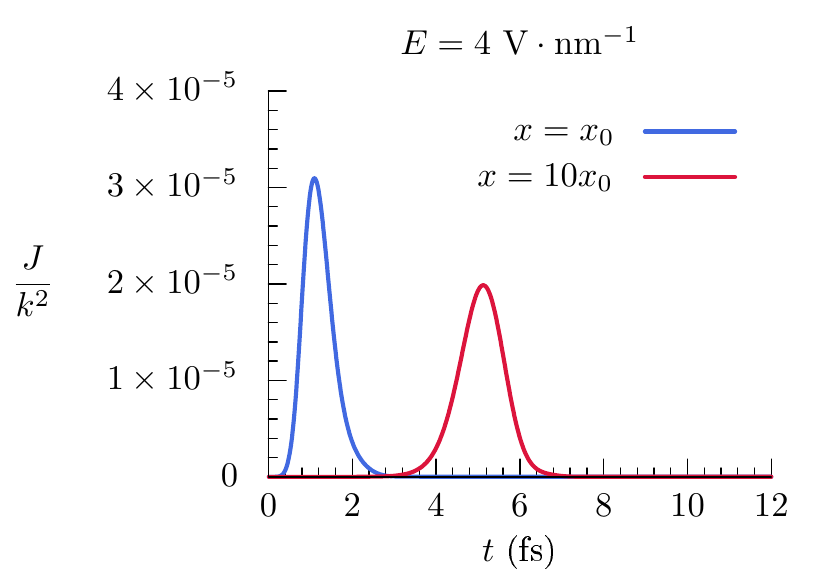}{\scriptsize({\bf e})}&
    \includegraphics[width=7cm]{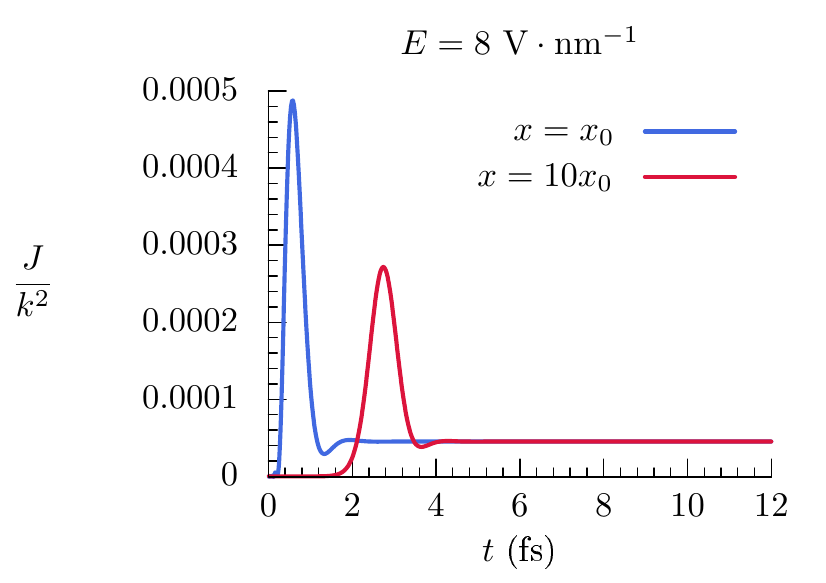}{\scriptsize({\bf f})}
  \end{tabular}
  \caption{The density ({\bf a}),({\bf b}), current ({\bf c}),({\bf d}) and integrated current ({\bf e}),({\bf f}) as a function of time at $x=x_0\equiv\frac{2U-k_{\mathrm F}^2}{2E}$ and $x=10x_0$. We have taken $U=9\ \mathrm{eV}$ and $k^2/2=k_{\mathrm F}^2/2\equiv E_{\mathrm F}=4.5\ \mathrm{eV}$. In ({\bf a}),({\bf c}),({\bf e}), the field is $E=4\ \mathrm{V}\cdot \mathrm{nm}^{-1}$ and $x_0\approx1.1\ \mathrm{nm}$. In ({\bf b}),({\bf d}),({\bf f}), $E=8\ \mathrm{V}\cdot\mathrm{nm}^{-1}$ and $x_0\approx0.56\ \mathrm{nm}$. In ({\bf a}),({\bf c}),({\bf e}), the plots seem to indicate that the curves converge to 0, but they actually tend to small finite values.}
  \label{current_density}
\end{figure}

In Fig.\-~\ref{current_density} we have plotted the density $|\psi(x,t)|^2$, current
\begin{equation}
  j_k(x,t):=i(\psi\partial_x\psi^*-\psi^*\partial_x\psi)
\end{equation}
and integrated current (the current integrated over the supply function at 0 temperature)
\begin{equation}
  J_{k_{\mathrm F}}(x,t):=\int_0^{k_{\mathrm F}}dk\ j_k(x,t)
\end{equation}
as a function of time at two different values of $x$: $x_0:=\frac{2U-k_{\mathrm F}^2}{2E}$ and $10x_0$ ($x_0$ is the point at which $V(x_0)=\frac{k_{\mathrm F}^2}2$), and at two different values of $E$: $4$ and $8\ \mathrm{V}\cdot\mathrm{nm}^{-1}$. We have normalized the current $j$ by $2k$, which is the current of the incoming wave $e^{ikx}$, and the integrated current $J$ by $k_{\mathrm F}^2$, which is the current of the incoming wave integrated over the supply function. We find that there is a transient regime that lasts a few femtoseconds before the system stabilizes to the FN value. Note that the approach to the FN regime has some ripples, which come from the imaginary parts of the poles in the $p$-plane (see Fig.\-~\ref{fig:contour}). There is a delay before the signal reaches $x_0$, and between $x_0$ and $10x_0$. As expected, the asymptotic value of the current is independent of $x$. Note that the current and density depend strongly on the field $E$.
\bigskip

\noindent{\bf Remark}: While the time scale of the approach to the FN solution is clearly of order of femtoseconds we have not attempted to compute a ``tunneling time''. This is, as is well known, a tricky business, with many possible definitions, see, e.g. \cite{LM94,LK15}. Defining such a time in terms of the approach of the initial state to some steady state was investigated by McDonald et al.\-~\cite{MOe13}. Pfeifer and Fr\"ohlich\-~\cite{PF95} have computed rigorous bounds on the lifetimes of spatially confined states.

\section{Possible generalizations}  

\subsection{Initial conditions}
As is shown by the computation described above, the long time asymptotic behavior of the wave function is independent of the initial reflected and transmitted waves. That is, the initial condition $\psi(x,0)=\psi^{(\mathrm I)}(x,0)=\Theta(-x)e^{ikx}$ leads to the same asymptotic formula:
\begin{equation}
  \psi(x,t)\sim e^{-i\frac12 k^2t}\psi_E(x).
\end{equation}
Indeed, the reflected and transmitted initial conditions do not actually give rise to any poles on the imaginary axis and therefore their contributions decay in time.
\bigskip

This leaves open the possibility to consider much more general initial conditions than\-~(\ref{psi_sum}): one can change the coefficients of the reflected and transmitted waves, add such waves with different wave vectors, or remove them altogether, without changing the asymptotic formula. Only the incoming wave $e^{ikx}$ affects it. In addition, one can add any square-integrable function to the initial condition without changing the long-time behavior. This is a consequence of the RAGE theorem \cite{Ru69,AG73,En78}, which states that whenever the Hamiltonian has absolutely continuous spectrum (as is the case here), the solution of the Schr\"odinger equation with a square-integrable initial condition vanishes point-wise as $t\to\infty$.
\bigskip

With this fact in mind, one can make an easy argument why the asymptotic behavior of the wave function must coincide with the stationary solution. Indeed, once we drop the initial reflected and transmitted waves,
 the Laplace transform of the wave function is of the form
\begin{equation}
  \hat\psi_p(x)=\frac{-2i}{-2ip+k^2}\left(\left(e^{ikx}+Re^{-\sqrt{-2ip}x}\right)\Theta(-x)+T\varphi_p(x)\Theta(x)\right)
\end{equation}
in which $\varphi_p$ is the solution of
\begin{equation}
  \left(-\frac12\partial_x^2+V(x)-ip\right)\varphi_p(x)=0
  ,\quad
  x>0
  .
  \label{eqvarphi}
\end{equation}
When $p=-ik^2/2$, (\ref{eqvarphi}) coincides with the equation\-~(\ref{eqpsiE}) for $\psi_E$. Assuming that $R$ and $T$ do not introduce any new poles on the imaginary axis (as we showed is the case), this implies that $\psi(x,t)$ converges to $e^{-i\frac12k^2t}\psi_E(x)$ as $t\to\infty$.
\bigskip

\subsection{Potentials}
The exact form of the potential $V(x)=U-Ex$ was not really used in much of the computation above, so it can be carried out in very much the same way for many other $V(x)$. For instance, one could round off the triangular barrier as occurs in the Schottky effect\-~\cite{Fo08}. We could also consider a square barrier. The only real constraint on the potential is that it does not introduce bound states. To make this into a precise statement, one would also have to put constraints on the regularity and asymptotic properties of $V$, which we will not do here.
\bigskip

This leaves open the possibility of studying trains of pulses, in which the field is turned on and off repeatedly. The regime in which the field is off corresponds to a potential $V(x)=U\Theta(x)$, which can be studied using the method described above. Provided the time between the pulses is long enough, the system would stabilize to the stationary state in the time between each field switching.
\bigskip

\subsection{Time-dependent fields}
It would be very interesting to consider a similar question in the case of an oscillating laser field $E=e_0\cos(\omega t)$. The stationary state of this problem was studied by Faisal et al.\-~\cite{FKS05}, and we are currently working on showing that the solutions of the initial value problem converge to this solution, and studying the short-time behavior.
\bigskip

\hfil{\bf Acknowledgements}\par\penalty10000
\noindent This material is based upon work supported by the AFOSR under the award number FA9500-16-1-0037. OC was partially supported by the NSF-DMS grant 1515755. IJ was partially supported by the NSF-DMS grant 1128155. JLL thanks Kevin Jensen and Don Shiffler for useful discussions and the IAS for hospitality during part of this work.


\begin{thebibliography}{WWW99}

\bibitem[AG73]{AG73}W.O. Amrein, V. Georgescu - {\it On the characterization of bound states and scattering states in quantum mechanics}, Helvetica Physica Acta, volume~\-46, issue~\-5, pages~\-635-658, 1973,\par\penalty10000
doi:{\tt\color{blue}\href{http://dx.doi.org/10.5169/seals-114499}{10.5169/seals-114499}}.\par\medskip
 
\bibitem[En78]{En78}V. Enss - {\it Asymptotic completeness for quantum mechanical potential scattering}, Communications in Mathematical Physics, volume~\-61, issue~\-3, pages~\-285-291, 1978,\par\penalty10000
doi:{\tt\color{blue}\href{http://dx.doi.org/10.1007/BF01940771}{10.1007/BF01940771}}.\par\medskip
 
\bibitem[FKS05]{FKS05}F.H.M. Faisal, J.Z. Kami\'nski, E. Saczuk - {\it Photoemission and high-order harmonic generation from solid surfaces in intense laser fields}, Physical Review A, volume~\-72, issue~\-2, number~\-023412, 2005,\par\penalty10000
doi:{\tt\color{blue}\href{http://dx.doi.org/10.1103/PhysRevA.72.023412}{10.1103/PhysRevA.72.023412}}.\par\medskip
 
\bibitem[Fo08]{Fo08}R.G. Forbes - {\it On the need for a tunneling pre-factor in Fowler–Nordheim tunneling theory}, Journal of Applied Physics, volume~\-103, issue~\-11, number~\-114911, 2008,\par\penalty10000
doi:{\tt\color{blue}\href{http://dx.doi.org/10.1063/1.2937077}{10.1063/1.2937077}}.\par\medskip
 
\bibitem[Fo08b]{Fo08b}R.G. Forbes - {\it Physics of generalized Fowler-Nordheim-type equations}, Journal of Vacuum Science and Technology B, volume~\-26, issue~\-2, pages~\-788-793, 2008,\par\penalty10000
doi:{\tt\color{blue}\href{http://dx.doi.org/10.1116/1.2827505}{10.1116/1.2827505}}.\par\medskip
 
\bibitem[Fo16]{Fo16}R.G. Forbes - {\it Field Electron Emission Theory}, Proceedings of Young Researchers in Vacuum Micro/Nano Electronics, IEEE, 2016,\par\penalty10000
doi:{\tt\color{blue}\href{http://dx.doi.org/10.1109/VMNEYR.2016.7880403}{10.1109/VMNEYR.2016.7880403}}, arxiv:{\tt\color{blue}\href{http://arxiv.org/abs/1801.08251}{1801.08251}}.\par\medskip
 
\bibitem[FN28]{FN28}R.H. Fowler, L. Nordheim - {\it Electron emission in intense electric fields}, Proceedings of the Royal Society of London A, volume~\-119, issue~\-781, pages~\-173-181, 1928,\par\penalty10000
doi:{\tt\color{blue}\href{http://dx.doi.org/10.1098/rspa.1928.0091}{10.1098/rspa.1928.0091}}.\par\medskip
 
\bibitem[Je03]{Je03}K.L. Jensen - {\it Electron emission theory and its application: Fowler-Nordheim equation and beyond}, Journal of Vacuum Science and Technology B: Microelectronnics and Nanometer Structures Processinf, Measurement and Phenomena, volume~\-21, issue~\-4, number~\-1528, 2003,\par\penalty10000
doi:{\tt\color{blue}\href{http://dx.doi.org/10.1116/1.1573664}{10.1116/1.1573664}}.\par\medskip
 
\bibitem[Je17]{Je17}K.L. Jensen - {\it Introduction to the Physics of Electron Emission}, Wiley, 2017.\par\medskip
 
\bibitem[LM94]{LM94}R. Landauer, T. Martin - {\it Barrier interaction time in tunneling}, Reviews of Modern Physics, volume~\-66, issue~\-1, pages~\-217-228, 1994,\par\penalty10000
doi:{\tt\color{blue}\href{http://dx.doi.org/10.1103/RevModPhys.66.217}{10.1103/RevModPhys.66.217}}.\par\medskip
 
\bibitem[LK15]{LK15}A.S. Landsman, U. Keller - {\it Attosecond science and the tunnelling time problem}, Physics Reports, volume~\-547, pages~\-1-24, 2015,\par\penalty10000
doi:{\tt\color{blue}\href{http://dx.doi.org/10.1016/j.physrep.2014.09.002}{10.1016/j.physrep.2014.09.002}}.\par\medskip
 
\bibitem[MOe13]{MOe13}C.R. McDonald, G. Orlando, G. Vampa, T. Brabec - {\it Tunnel Ionization Dynamics of Bound Systems in Laser Fields: How Long Does It Take for a Bound Electron to Tunnel?}, Physical Review Letters, volume~\-111, issue~\-9, number~\-090405, 2013,\par\penalty10000
doi:{\tt\color{blue}\href{http://dx.doi.org/10.1103/PhysRevLett.111.090405}{10.1103/PhysRevLett.111.090405}}.\par\medskip
 
\bibitem[PF95]{PF95}P. Pfeifer, J. Fröhlich - {\it Generalized time-energy uncertainty relations and bounds on lifetimes of resonances}, Reviews of Modern Physics, volume~\-67, issue~\-4, pages~\-759-779, 1995,\par\penalty10000
doi:{\tt\color{blue}\href{http://dx.doi.org/10.1103/RevModPhys.67.759}{10.1103/RevModPhys.67.759}}.\par\medskip
 
\bibitem[Ro11]{Ro11}A. Rokhlenko - {\it Strong field electron emission and the Fowler-Nordheim-Schottky theory}, Journal of Physics A: Mathematical and Theoretical, volume~\-44, issue~\-5, pages~\-1-10, 2011,\par\penalty10000
doi:{\tt\color{blue}\href{http://dx.doi.org/10.1088/1751-8113/44/5/055302}{10.1088/1751-8113/44/5/055302}}.\par\medskip
 
\bibitem[Ru69]{Ru69}D. Ruelle - {\it A remark on bound states in potential-scattering theory}, Il Nuovo Cimento, volume~\-61, issue~\-4, pages~\-655-662, 1969,\par\penalty10000
doi:{\tt\color{blue}\href{http://dx.doi.org/10.1007/BF02819607}{10.1007/BF02819607}}.\par\medskip
 
\bibitem[YGR11]{YGR11}S.V. Yalunin, M. Gulde, C. Ropers - {\it Strong-field photoemission from surfaces: Theoretical approaches}, Physical Review B, volume~\-84, issue~\-19, number~\-195426, 2011,\par\penalty10000
doi:{\tt\color{blue}\href{http://dx.doi.org/10.1103/PhysRevB.84.195426}{10.1103/PhysRevB.84.195426}}.\par\medskip
 
\bibitem[ZL16]{ZL16}P. Zhang, Y.Y. Lau - {\it Ultrafast strong-field photoelectron emission from biased metal surfaces: exact solution to time-dependent Schr\"odinger Equation}, Scientific Reports, volume~\-6, number~\-19894, 2016,\par\penalty10000
doi:{\tt\color{blue}\href{http://dx.doi.org/10.1038/srep19894}{10.1038/srep19894}}.\par\medskip
 
\end{thebibliography}
\end{document}